# Luminous Flame Height Correlation Based on Fuel Mass Flow for a Laminar to Transition-to-Turbulent Regime Diffusion Flame

M. De la Cruz-Ávila [a*], J. E. De León-Ruiz [b], I. Carvajal-Mariscal [b], G. Polupan [b] and L. Di G. Sigalotti [c]

[a*] Universidad Nacional Autónoma de México, Instituto de Ingeniería, Ciudad Universitaria, 04510 México City, MÉXICO.
[b] Instituto Politécnico Nacional, ESIME UPALM, Av. IPN s/n, 07738 México City, MÉXICO.
[c] Universidad Autónoma Metropolitana-Azcapotzalco (UAM-A), Departamento de Ciencias Básicas, Av. San Pablo 180, 02200 México City, MÉXICO.
*Correspondence: mauriciodlca1@gmail.com

**ABSTRACT**

This paper presents a flame-height correlation for laminar to transition-to-turbulent regime diffusion flames. Flame-height measurements are obtained by means of numerical and experimental studies in which three high definition cameras were employed to take frontal, lateral and 45° angled images simultaneously. The images were analysed using an image-processing algorithm to determine the flame-height through indirect measurement. To locate an overall chemical-flame-length, numerical simulations were conducted with the unsteady Reynolds-Averaged Navier-Stokes technique. The eddy-dissipation model was also implemented to calculate chemical reaction rate. The experiments show that this proposed correlation has an adjustment variation of luminous flame-height for the laminar regime of 16.9%, which indicates that, without the use of the intermittent buoyant flame-height correlation, it globally best represents the flame-height in this regime. For the laminar and transition-to-turbulence regime the adjustment variations are 5.54% compared to the most accepted flame-height correlations, thus providing an acceptably good fitting. The numerical results show that the proposed range for the chemical-flame-length is located between the luminous and flickering flame zone compared to the experimental flame images. These results agree with the chemical length zone reported in the literature. Therefore, the correlation can be used for laminar and transition-to-turbulent combustion regimes.

## 1 INTRODUCTION

The design of most combustion chambers, as in the applied technology of in-situ steam generator prototypes [1,2], requires data on several fundamental parameters, such as flame stabilization, emission characteristics, auto ignition, combustion dynamics, etc. These parameters are influenced by the fuel properties and the oxidant mixture [3,4] as, for example, the flame velocity, the ignition delay, the minimum ignition energy and the flammability limits [5]. In this area, one of the primary energy sources in commercial applications and domestic use is the propane [3,6,7], which is the principal component of liquefied petroleum gas (LPG). Some authors have focused on combined numerical [8–11] and experimental studies of propane-air mixtures diffusion flames within combustion chambers [12–15]. In these experimental investigations of inverse diffusion flames, the analysis of the flame length is of fundamental importance since it provides information on the length of the flame front, the length of the luminescence structure, the stoichiometric mixing zone and the expansion flame radius in order to obtain the different velocities at which the zone of turbulent diffusion flame develops. It is precisely from these characteristics that the positions where the transition from laminar to turbulent flame zone occurs can be accurately estimated.

Many definitions and techniques for measuring flame lengths can be found in the literature. However, none of these definitions is preferred to the others. Therefore, care must be exercised in comparing results of different authors and in the application of correlation formulae. Common definitions of flame length include visual determinations as obtained by averaging several individual instantaneous visible flame lengths from photographic records. Such length estimates are obtained by measuring the axial location of the average peak centre-line temperature and the axial location where the mean mixture fraction on the flame axis is the stoichiometric value.

For many past decades there has been strong interest to explain and predict the turbulent jet flame lengths. The earliest, and perhaps the most representative, studies are those of Burke and Schumann [16], Hawthorne *et al*. [17], Wohl *et al*. [18], Yagi [19], Hottel [20], Powell [21], Steward [22], Podymov and Chuchalin [23], Yarin [24], Steward [25], Hill *et al*. [26], Brzustowski [27], Baev *et al*. [28], Suris *et al*. [29] and Becker and Yamazaki [30] to quote a few authors. A major review on this subject is given by Becker and Liang [31]. Already in the 80s many theories and correlations were consolidated and plenty of data were collected [32–37]. More recent work in this line is provided by Delichatsios [38], Reshetnikov and Frolov [39] and Blake and McDonald [40]. Since then, Heskestad [41,42] is one of the few authors who has continued to develop new correlations. These correlations have been tested for fuel pool fire, premixed flames, laminar diffusion flames and co-flow diffusion flames. However, for recently inverse diffusion flames or port array configurations, neither low nor high pressures have been tested due to the buoyant-flickering nature of the flame. Because of this, Zukoski *et al*. [43] proposed for the first time a correlation using more extensively the Froude number to determine a general flame length, based on heat release, gravity effects and density gradients, which was later on adopted by Delichatsios [38]. In general, flame heights have been measured from videotape recordings and by optical averaged techniques, whose output although adequate, was not very accurate.

In general, visible flame lengths tend to be larger than those based on temperature or concentration measurements. For example, Turns and Bandaru [44] report temperature-based flame lengths which were approximately from 65 to 80 percent larger than the time-averaged visible flame lengths, depending on the fuel type. Similarly, Imamura *et al*. [45] argued that flame heights with a comparatively high initial discharging velocity increase with the rate of heat release. On the other hand, Bradley *et al.* [46] expanded this informational data by including six new regimes of fuel flow injection, based on the rate of heat release.

Knowledge of the visible length of turbulent diffusion flames is important for both practical and conceptual reasons [40]. In theory, the visible length of a non-premixed flame is an important indicator of the overall fuel-oxidant mixing process, since the flame length is proportional to the axial distance required to dilute the fuel mixture fraction to its stoichiometric value [47]. Flame height measurements have been used to test flame structure models and calculate residence times of ash particles [48]. In the industry, the flame length is of particular interest because, appropriate separation distances between the equipment and between the injection and the burner walls have to be specified for a given flame length [10].

The most commonly accepted definition of the height of the flame is given by the distance from the burner to the position in the centre line where the fuel and the oxidant are in stoichiometric proportions [48]. A more precise and less subjective methodology based on measurements of species along the flame axis is proposed. This methodology is also suitable for scenarios where observations of the luminous structure height of the flame are not possible [49]. The flame height is determined using a methodology called *Chemical Flame Height*, which differs from the luminous flame height determined from observations or by direct visualization [49]. The concept of chemical flame height is not a novel idea. It was proposed for the first time by Hawthorne *et al*. [17] and is defined by the distance to the point where 99% of the combustion is completed. On the other hand, Hottel [50] defined the height of the chemical flame as the axial location in which the CO over $CO_2$ ratio is 0.15, without giving a justification for this choice. More recently, Wade and Gore [51] defined the chemical height of the flame as the location at which the molar fraction of the fuel, $X_{fuel}$, along the fuel axis, falls to 0.0005. The justification for this value was reported as the lower limit for the hydrocarbon gas analyser [49]. The difficulty of this method lies in the two-dimensional distribution of the species since the hydrocarbon gas analyser is only a point measure. In addition, an intrusive measurement would perturb the flow field. The height of the flames is usually determined by direct observations of the flame brightness. The most common method used to measure the flame height is by visualizing the luminous shape because the stoichiometric conditions occur on the oxygen side of a reaction zone caused by $CO_2$. This is why the main focus here is to measure the height of the flame ($h_{Flame}$). To this aim a thorough experimental design was implemented, accompanied with an extensive statistical analysis and an accurate image processing methodology. Within this framework, the flame image analysis was carried out by means of an image-recognition and pixel-quantification algorithm of the flame contour shape, in order to locate the beginning of the laminar combustion process (the blue cone) and its development up to the turbulent flame zone (the high intense luminous flame).Therefore, with the knowledge of these characteristics, the proposed correlation can be applied to design and develop prototypes applied to enhance combustion chambers and improve, for example, direct-vaporization for in-situ steam-generator technology to name few.

## 2 EXPERIMENTAL SETUP

The experimental rig is shown in Fig. 1. It utilizes a gas-burner, central-peripheral fuel injection system to study the behaviour and define the structure of a laminar to transition-to-turbulent non-confined diffusion flame (Fig. 1a). The proposed setup employs as reacting species LPG as fuel and air as the oxidizing agent. For this experimental study, the LPG is composed approximately of 60% propane and 40% butane. This composition has thermo-chemical and reaction properties very similar to those reported by Mishra and Rahman [7].The experimental evaluation was designed to determine the flame height for six different fuel flows: (i) 350 cc/min, (ii) 650 cc/min, (iii) 950 cc/min, (iv) 1200 cc/min, (v) 1500 cc/min and (vi) 1800 cc/min as listed in Table 1. The fuel injection system makes use of a four-nozzle port array instead of the whole 4-Lug-Bolt setup. This arrangement consists of four 0.8 mm peripheral nozzles in a 4 x 16.94 mm radial distribution configuration for a 25.4 mm diameter gas-burner, as depicted in Fig. 1b.The 4 mm central nozzle was not employed here and was left aside to complement an ongoing research project involving the Lug-Bolt configuration. The proposed diameters of the gas-burner were calculated in order to maintain a stoichiometric ratio between the fuel and the oxidizing agent of 15.53 (see Table 1).

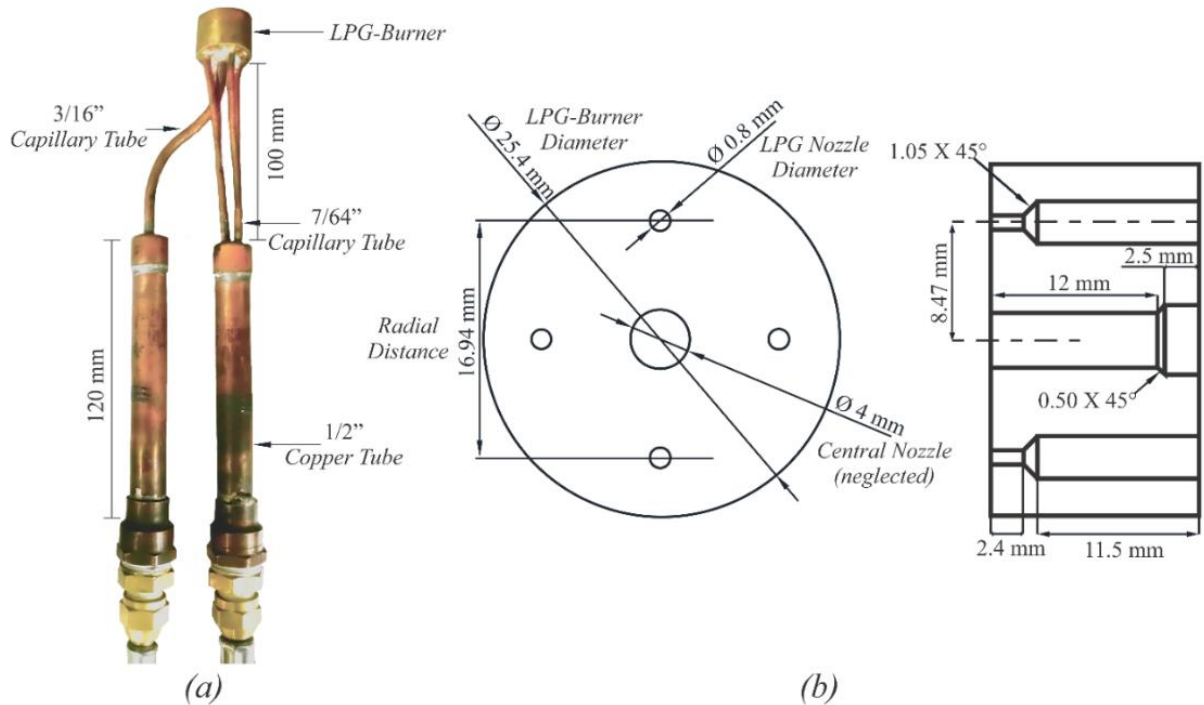

Fig. 1: Central-peripheral fuel injection system: (a) Distribution setup; (b) gas-burner configuration.

Table 1: LPG-air thermodynamic properties at 293.15 K and 0.7647 atm.

| Case Studies | a | b | c | d | e | f |
|---|---|---|---|---|---|---|
| LPG Injection Velocity [m/s] | 2.902 | 5.389 | 7.875 | 9.948 | 12.434 | 14.921 |
| LPG Mass Flow [kg/s] | $8.91 \times 10^{-6}$ | $1.65 \times 10^{-5}$ | $2.41 \times 10^{-5}$ | $3.05 \times 10^{-5}$ | $3.81 \times 10^{-5}$ | $4.58 \times 10^{-5}$ |
| LPG Volumetric Flow [cc/min] | 350.1 | 650.1 | 950.0 | 1200.1 | 1500.0 | 1800.0 |
| Injection Re GLP | 498.1 | 925 | 1351.6 | 1707.4 | 2134 | 2561 |
| Stoichiometric Air Mass Flow needed [kg/s] | $1.67 \times 10^{-4}$ | $3.10 \times 10^{-4}$ | $4.54 \times 10^{-4}$ | $5.73 \times 10^{-4}$ | $7.17 \times 10^{-4}$ | $8.61 \times 10^{-4}$ |
| Mass Air–Fuel Ratio $(A/F)_{stoich}$ | 15.53 | | | | | |
| Stoichiometric Mixture Fraction fs | 0.060477 | | | | | |

The experimental rig was designed to reduce the pressure drop in the fuel side with symmetric proportions, as shown in Figs. 1a and 2a. The instrumentation employed for this experiment is described as follows. The fuel supply system comprises a Parker pressure regulation valve Mod. N400S for robust control of the volumetric flow and a Dwyer flowmeter Mod. RMA 14 (4% F.S.) for fine adjustment and measuring. To measure the flame height, three high-definition cameras, circularly arranged at 1 m from the gas-burner, were employed to take frontal, lateral and 45° angled images of the flame, as seen in Fig. 2b. This array was complemented with a rig of interconnected linear electric actuators, fixed on the top of each camera, in order to trigger the shutter buttons simultaneously. Subsequently, the images obtained were analysed using MATLAB with the aid of an image-processing algorithm to determine the flame height through indirect measurement of the image pixels. In addition to this, the flame temperature was determined through 6 s measurements, employing a Heraeus pyrometer Mod. DT-400 (1% F.S.) with a tungsten-rhenium thermocouple probe and a Fluke-Ti55FT thermal-imaging camera which was located in the workbench frontal-plane at 1 m from the gas-burner to avoid emissivity errors [52–54].

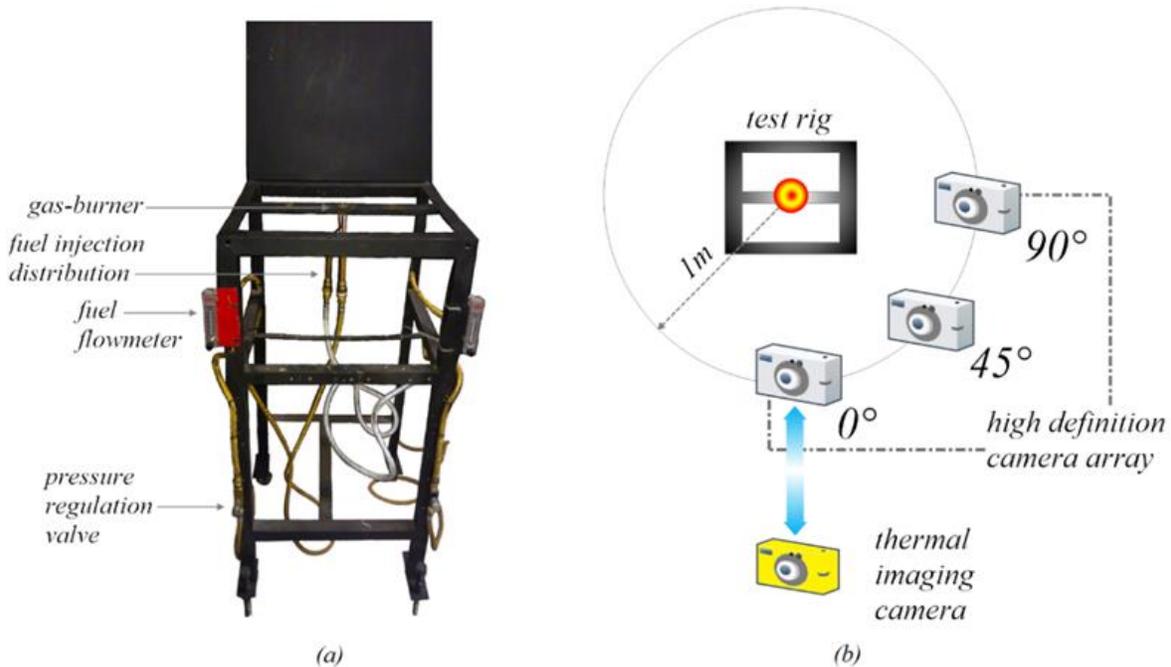

Fig. 2: Experimental setup: (a) Test rig; (b) camera-visualization array.

Finally, in order to keep the measurement errors within 10% and to ensure that such a measurement lies within a 95% confidence interval (CI), it was determined that a minimum of 96 samples was required. For this particular evaluation, this number was doubled and 32 treatments for every flow condition were conducted. The experiment proposed employed a completely randomized block design to minimize the error propagation due to both round-off / forward-carry, as well as due to the instrumentation. The factors and levels of the experimental design are summarized in Table 2.

Table 2: Factors and levels of the experimental design.

| Factors | Levels |
|---|---|
| Camera Direction | Lateral, Frontal, Angled |
| Volumetric Flow [cc/min] | 350, 650, 950, 1200, 1500, 1800 |

## 3  NUMERICAL SIMULATIONS
### 3.1  CASE STUDY

The numerical model considers an unconfined LPG-air diffusion flame with the same dimensions of the burner configuration shown in Fig. 1b. The nozzles are arranged in a radial distribution to maintain the air entrainment to the stoichiometric relation. The virtual combustion domain has a diameter $d=150$ mm and a total length $L=500$ mm. The injection nozzles have the exact geometrical array of the 4 x 16.94 mm radial distribution as shown in Fig.3a. The numerical study matrix is performed for the six different fluid flows with thermodynamic conditions as listed in Table 1. The analysis is focused in two precise zones, which correspond to those where the temperature and the $CO_2$ mol fraction are fully developed. These zones are intended to cover the overlapped region between the temperature and $CO_2$ mol fraction layers, respectively.

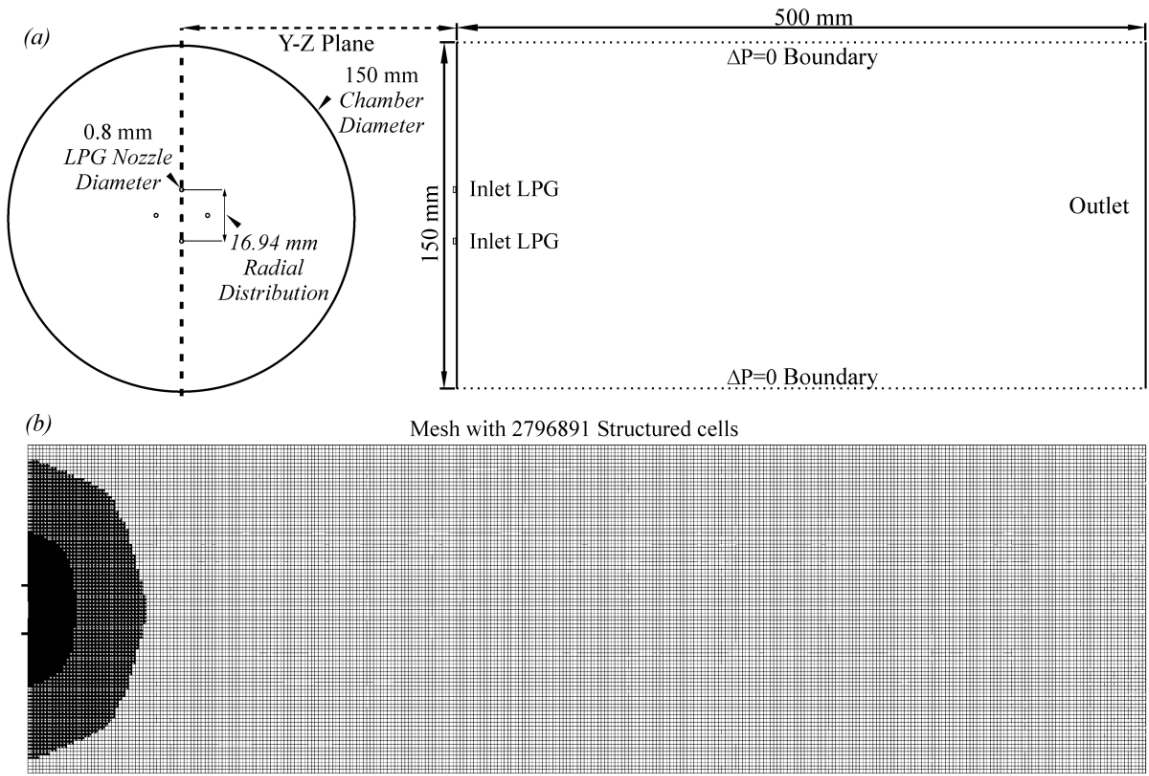

Fig.3: (a) Model geometry; (b) mesh details.

## 3.2 NUMERICAL DETAILS

The numerical simulations are based on a combination of the advancing-front meshing [55,56] and structured cell methods. One of the advantages offered by the advancing-front method over commonly structured grids is the facilitating "*tessellation*" process for geometrically complicated domains, allowing the mesh density to be adapted to the geometry. These merging methods result in a growth of thin layers and cells from the bottom wall of the combustion chamber to the final domain edge, thereby allowing the implementation of high-order discretization schemes. In addition to these meshing methods, an adaptive time-stepping was implemented to ensure the correct time-marching in complex simulations as occurs in the combustion process, where both high speeds and sudden energy releases are present. Table 3 lists the resulting time-steps as calculated using equation (1), which must be small enough to solve time-dependent features and to ensure convergence within a finite number of maximum iterations.

$$\Delta t \approx \frac{Typical\ cell\ size}{Characteristic\ flow\ velocity} \tag{1}$$

Table 3: Time stepping characteristics for the temporal discretization of the combustion process.

| Time Stepping Method | *a* | *b* | *c* | *d* | *e* | *f* |
|---|---|---|---|---|---|---|
| *Fixed* | | | $1 \times 10^{-4}$ | | | |
| *Adaptive* | $1 \times 10^{-2}$ $> \Delta t >$ $1 \times 10^{-4}$ | $7 \times 10^{-3}$ $> \Delta t >$ $1 \times 10^{-4}$ | $4 \times 10^{-3}$ $> \Delta t >$ $1 \times 10^{-4}$ | $1 \times 10^{-3}$ $> \Delta t >$ $1 \times 10^{-4}$ | $7 \times 10^{-4}$ $> \Delta t >$ $1 \times 10^{-5}$ | $4 \times 10^{-4}$ $> \Delta t >$ $1 \times 10^{-5}$ |

The simulations considered a mass flow injection through each nozzle (see Table 1). A zero pressure gradient condition is implemented at the outlet boundary and the gas discharges are allowed to occur at reduced atmospheric conditions of 0.7647atm and temperature of 293K. Furthermore, the Monotonic Upwind Scheme for Conservation Laws (MUSCL) scheme [57] was used for evaluating the convective and viscous terms, which provides a highly accurate numerical solution for the system. Finally, a sensitivity analysis was developed in order to obtain numerical results independent of the mesh resolution with deviations less than 5% as shown in Ref. [58]. Figure 3b shows the resulting mesh for the case of 2796891 structured cells. All numerical simulations were performed using an academic license Ansys Fluent v13 software.

## 3.3 GOVERNING EQUATIONS

The equations to be solved are the conservation laws of mass, momentum, energy and chemical species. A density-weighted averaging (Favre-averaging) denoted by "~" is considered, along a time average denoted by "¯". These terms are used in the transport equations to model the flow movement and the heat exchange. The Favre-averaged continuity, momentum, energy and species equations are as follows:

***Mass conservation:***

$$\frac{\partial \bar{\rho}}{\partial t} + \nabla \cdot (\bar{\rho} \widetilde{\mathbf{u}}) = 0. \tag{2}$$

*Momentum:*

$$\bar{\rho}\frac{\partial \tilde{\mathbf{u}}}{\partial t} + \bar{\rho}\tilde{\mathbf{u}} \cdot \nabla \tilde{\mathbf{u}} = -(\nabla \bar{p}) + \nabla \cdot \bar{\tau} + \bar{\rho}\sum_{i=1}^{N} \widetilde{Y_i \mathbf{f}_i} - \nabla \cdot (\bar{\rho}\widetilde{\mathbf{u}'\mathbf{u}'}). \quad (3)$$

*Conservation of Energy:*

$$\bar{\rho}\frac{\partial \tilde{e}}{\partial t} + \bar{\rho}\tilde{\mathbf{u}} \cdot \nabla \tilde{e} = -\nabla \cdot \tilde{\mathbf{q}} - \overline{p\nabla \cdot \mathbf{u}} + \overline{\tau:\nabla \mathbf{u}} + \bar{\rho}\sum_{i=1}^{N} \widetilde{Y_i \mathbf{f}_i \cdot \mathbf{V}_i} - \nabla \cdot (\bar{\rho}\widetilde{\mathbf{u}'e'}). \quad (4)$$

*Species:*

$$\bar{\rho}\frac{\partial \tilde{Y}_i}{\partial t} + \bar{\rho}\tilde{\mathbf{u}} \cdot \nabla \tilde{Y}_i = \nabla \cdot \left(-\bar{\rho}\widetilde{\mathbf{V}_i Y_i}\right) + \bar{\omega}_i - \nabla \cdot (\bar{\rho}\widetilde{\mathbf{u}'Y_i'}), \quad i = 1, \dots, N, \quad (5)$$

where $\bar{\rho}\widetilde{\mathbf{u}'\mathbf{u}'}$, $\bar{\rho}\widetilde{\mathbf{u}'Y_i'}$ and $\bar{\rho}\widetilde{\mathbf{u}'e'}$ are the Reynolds stress tensor, the mass-weight density fluctuations, and the turbulent heat transfer vector, respectively. $\omega_i$ is the $i^{th}$ species production rate and $e$ is expressed as $e = \sum_{i=1}^{N} h_i Y_i - p/\rho$. For numerical solution these equations are written in Cartesian coordinates.

Equations (3), (4) and (5) demand additional mathematical expressions for terms like $\bar{\rho}\widetilde{\boldsymbol{\phi}'\mathbf{u}'}$. The closure of these equations requires modelling the Reynolds stress tensor, the turbulent heat flux and the mass-weight density fluctuations. Then, the Reynolds stress tensor is closed using the Realizable *k-ε* model [59]. The turbulent heat flux vector and the mass-weight density fluctuations are obtained by means of an analogy between momentum transfer and molecular diffusion. The Realizable *k-ε* turbulence model [59,60] has been validated experimentally for many reactive flows with satisfactory results [8,61–63]. This model is analogous to the standard *k-ε* model, except that $C_\mu$ is now handled as a variable, which improves the calculation of viscous effects and provides increased precision in combustion analysis.

### 3.4 COMBUSTION MODELLING

The species are introduced by means of their mass fractions $Y_i$ for $i=1$ to $N$, where $N$ specifies their number in the reactive mixture. The mass fractions, $Y_i$, are defined by the relation

$$Y_i = \frac{m_i}{m}, \quad (6)$$

where $m_i$ is the mass of species $i$ present in a given volume $V$ and $m$ is the total mass of gas in the volume. The energy production of LPG is established by the overall reaction

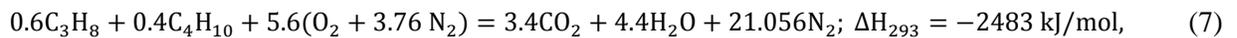

$$0.6 C_3H_8 + 0.4 C_4H_{10} + 5.6(O_2 + 3.76\, N_2) = 3.4 CO_2 + 4.4 H_2O + 21.056 N_2; \quad \Delta H_{293} = -2483 \text{ kJ/mol}, \quad (7)$$

which implies a gross simplification with the actual reaction mechanism and involves many free-radical chain reactions. Nevertheless, the main purpose of the present work is not to analyse the secondary chemical reactions. For this reason, a single-step irreversible chemical reaction was used to focus on the flow development

The turbulent chemical reaction rate is modelled with the aid of the Eddy Dissipation Model (EDM) [64], which is based on the infinitely fast chemistry hypothesis and assumes that the reaction rate is controlled by the turbulent mixing. The EDM generalized formulation has been proposed in order to take into account finite-rate chemistry effects. A stoichiometric relation describing chemical reactions of arbitrary complexity can be represented by the $r^{th}$ reaction equation [65]. The production net-rate of species $i$ due to reaction $r$, $R_{ir}$, is given by the smallest *limiting-value* of the two expressions

$$R_{i,r} = v'_{i,r} M_{w,i} A\rho \frac{\varepsilon}{k} \min_{\mathcal{R}} \left( \frac{Y_{\mathcal{R}}}{v'_{\mathcal{R},r} M_{w,\mathcal{R}}} \right), \tag{8}$$

$$R_{i,r} = v'_{i,r} M_{w,i} AB\rho \frac{\varepsilon}{k} \frac{\sum_P Y_P}{\sum_j^N v''_{j,r} M_{w,j}}, \tag{9}$$

which are based on reactants and products mass fractions, where $Y_P$ and $Y_R$ are the species mass fraction of products and reactants, respectively, while $A$ and $B$ are the Magnussen constants [64] for reactants (4.0) and products (0.5), respectively. Moreover, $M_{w,i}$ is the molecular weight for both reactants and products.

## 4 EXPERIMENTAL ANALYSIS
### 4.1 ANALYSIS OF VARIANCES

An analysis of variances, ANOVA, is performed to evaluate the effect of the factors on the output, i.e. on the measured flame height. The results of this analysis are summarized in Table 4. The analysis reveals that, among the factors involved, the fuel flow exerts a stronger effect, given a P-Value lower than 0.05, which states that within a 95% CI there is a significant statistical difference. The high mean square between groups indicates that a change on the magnitude of the fuel flow does affect significantly the height output, whilst the data fluctuation within each group is practically negligible. Conversely, it is seen that for the camera direction, a P-Value of 0.2781, which is much higher than 0.05, is obtained. This means that there is no a statistical significant difference between the samples. Albeit, a relatively high mean square between and within the groups, a low F-Ratio is obtained, which implies that even though the samples fluctuate, this is not significantly affecting the output. Nonetheless, this sum of squares entails that the position of the camera does induce slight variations of the output, which are mostly attributed to the fact that in some samples the skewness of the flame can be seen in one plane but not in the other. In particular, this occurs for the frontal-lateral ratio, with the flame height consequently differing between images. Meanwhile, within the groups, the height variation is primarily attributed to the occasional capture of flame kernels still attached to the flame body.

Table 4: ANOVA for each factor of the experiment.

|  | Source | Sum of Squares | Df | Mean Square | F-Ratio | P-Value |
|---|---|---|---|---|---|---|
| Fuel Flow | Between groups | 45679.88 | 5 | 9135.975 | 2617.41 | 0.0000 |
|  | Within groups | 1989.566 | 570 | 3.490467 |  |  |
|  | Total (Corr.) | 47669.44 | 575 |  |  |  |
| Camera Direction | Between groups | 212.487 | 2 | 106.2435 | 1.28 | 0.2781 |
|  | Within groups | 47456.96 | 573 | 82.82191 |  |  |
|  | Total (Corr.) | 47669.44 | 575 |  |  |  |

To complement this analysis, the average flame heights for each of the fuel flow analysed and for each of the three allowed camera directions are plotted in Fig. 4. It is seen that for the two fuel flows that are closer to the upper limit, the height difference between the camera directions is relatively negligible, with coefficients of variation of 2.25% for the 1500 cc/min flow and 0.21% for the 1800 cc/min flow. Meanwhile, the highest coefficient of variation occurs for the 650 cc/min flow with 5.29%, followed by the 950cc/min, 350cc/min and 1200 cc/min flows with 4.55%, 4.31% and 3.49%, respectively. These results reveal that by increasing the fuel flow, the resulting flame stabilizes, thus mitigating fluctuations due to height flickering.

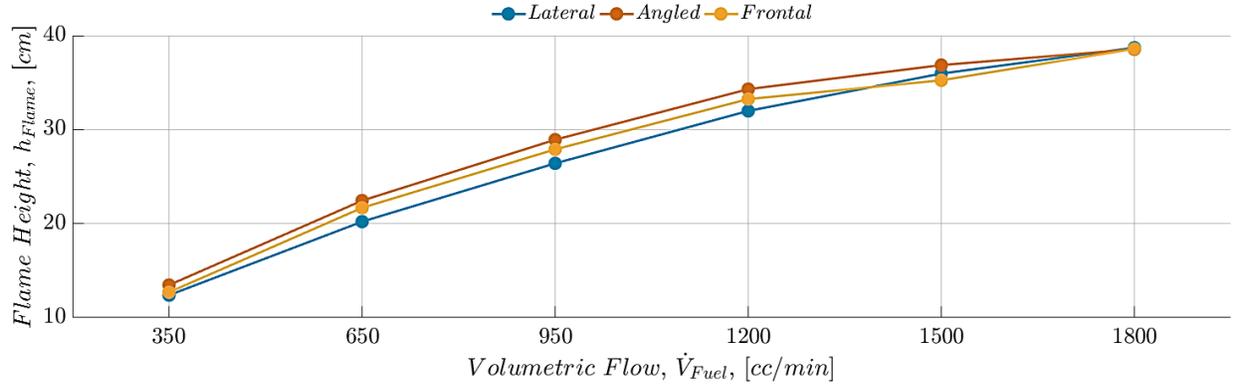

Fig. 4: Average flame height, $h_{Flame}$, as a function of the camera direction and the fuel flow.

### 4.2 FLAME TEMPERATURE ANALYSIS

If the combustion reaction is carried out under adiabatic conditions then $\delta Q=0$ at constant pressure, and the first law of thermodynamics yields $dH= 0$, where the energy released in the combustion process raises the thermal level of the reaction products even more, while the overall mass remains constant. Therefore, the burnt and unburnt gases, denoted by superscripts *b* and *u*, respectively, have the same specific enthalpy. The temperature reached by the products under these conditions is called the adiabatic flame temperature, $T_{adiab}$. Applying the first principle of thermodynamics and considering that a high percentage of the combustion processes take place at constant pressure, it can be verified that

$$\delta Q = dE + \delta W = de + pdV = dH, \qquad (10)$$

$$\delta Q = 0 \rightarrow dH = 0 \rightarrow H^b = H^u, \qquad (11)$$

where *Q* is the heat, *H* and *h* are the enthalpies, *W* is the work, *E* and *e* denote energy and *p* is the pressure. Furthermore, the molar enthalpies of the burnt and unburnt gases often differ because the amount of molecules usually changes in a chemical reaction. Thus,

$$h^u = \sum_{j=1}^{S} w_j^u h_j^u = \sum_{j=1}^{S} w_j^b h_j^b = h^b, \qquad (12)$$

and for constant pressure the relation holds

$$h_j^b = h_j^u + \int_{T_u}^{T_b} c_{p,j} dT. \qquad (13)$$

The adiabatic flame temperature can be obtained from the balance established between the enthalpy of both burnt and unburnt gases. For this process, it is necessary to know the composition of both the reactive and products mixtures in order to evaluate the adiabatic flame temperature $T_{adiab}$. The maximum adiabatic flame temperature is reached when the combustion process in adiabatic conditions corresponds to a complete combustion of a stoichiometric mixture. Under other conditions, the adiabatic flame temperature will have a lower value. The adiabatic flame temperature, $T_{adiab,}$ can be determined from equation (13), which corresponds to the temperature resulting after combustion provided that heat losses to the surroundings are negligible. The LPG adiabatic flame temperature, $LPG$-$T_{adiab}$, can be computed easily by means of a simple iteration method.

In commonly used combustion setups, the experimental area is so ventilated that the mixing process is ensured, with excess air in most cases. This causes a significant variation, with a consequent reduction of the flame temperature and with the corresponding measurement yielding a lower output compared to the one obtained by analytical modelling. For this particular instance, the resulting theoretical magnitude equals approximately 2398 K, which agrees with Silverman's [52] values. Conversely, it is seen in Fig. 5 that the measured experimental temperature of the flame presents a relatively constant behaviour, despite the changes on the fuel flow, with a coefficient of variation of 0.29% and an overall average of 2366 K. Additionally, the average temperature discrepancy between the theoretical and experimental magnitudes lies between 20 and 45 K, which is small compared to the actual temperature reached by the flame.

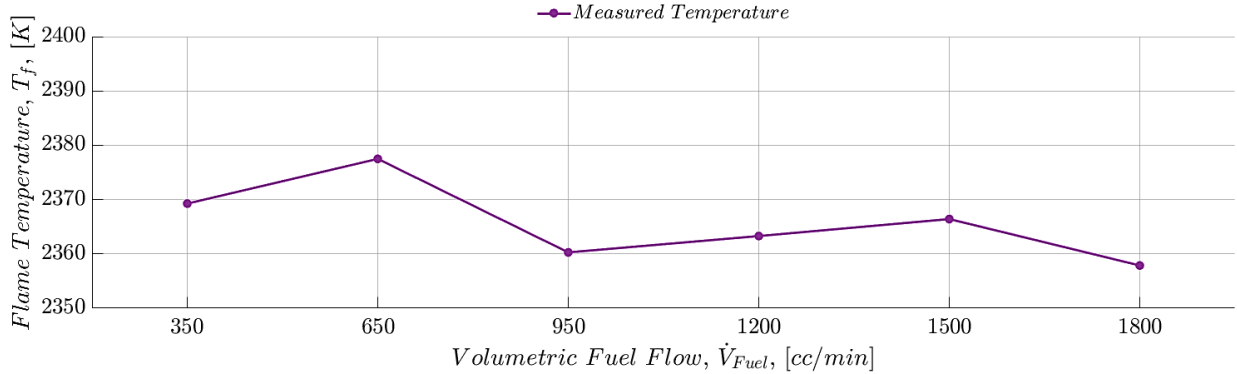

Fig. 5: Average measured flame temperature, $T_{Flame}$, as a function of the fuel flow.

It is also seen from Fig. 5 that the flame temperature measurements remain constant due to the well-controlled conditions (i.e., a properly ventilated experimental area), resulting in a near-the-most stoichiometric mixture. This mixture contains almost the exact amount of fuel and oxidizer such that air has no combustion products even at completed combustion. For all regimes, the fuel and oxidizer are consumed to form products, which means that the entrained air reaches the flame without any perturbation. This ideal mixture approximately yields the maximum flame temperature, as all the energy released from combustion is used to maintain the heat of the products.

### 4.3 FLAME HEIGHT ANALYSIS

It must be noticed that the luminous flame height is the length where all the dimensions related to the flame are included, namely the flame core height, the lift-off height as well as the dimensions implied by all mixing processes, instabilities and internal and external vortex structures zones, the chemical flame height, and the dimensions of the buoyant and flickering flame zone, the necking and the oscillation tip frequency. On the tip of this length one can locate the plumes, the $NO_x$, $SO_x$ and ashes generation zone, the cut-off kernels and more. To make a good recognition of the luminous flame heights, it is necessary to add the height provided by the intermittent flame correlation plus that given by the consistent height correlation.

From the height and temperature magnitudes obtained, as well as their corresponding increment rates, or lack thereof, it is inferred that a correlation exists between the studied factors; particularly the fuel flow, $\dot{V}_{Fuel}$, and the measured variables; specifically the flame height, $h_{Flame}$. Based on the data analysis, statistical modelling is applied and the equation obtained, which comes from a double reciprocal modelling. That is, *a correlation of the form $y = (a + b/x)^{-1}$ is proposed* for the luminous flame height, which is given by the relation

$$h_{Flame} = \left(0.012161 + \frac{22.8297}{\dot{V}_{Fuel}}\right)^{-1}. \tag{14}$$

The model proposed achieves a coefficient of determination, $R^2$, of 0.9984, which explains the percentage variability of the flame height based on the fitted model, a mean absolute error (MAE) of 0.0006165, as the average value of the residuals, and a standard error of estimate (SSE) of 0.0008616, describing the negligible deviation of the residuals. These values are considered more than acceptable in terms of the model adjustment and thus they could be used to construct prediction limits for new observations. Figure 6 shows the processed flame images and Table 5 lists the values of the overall average flame height for all studied cases.

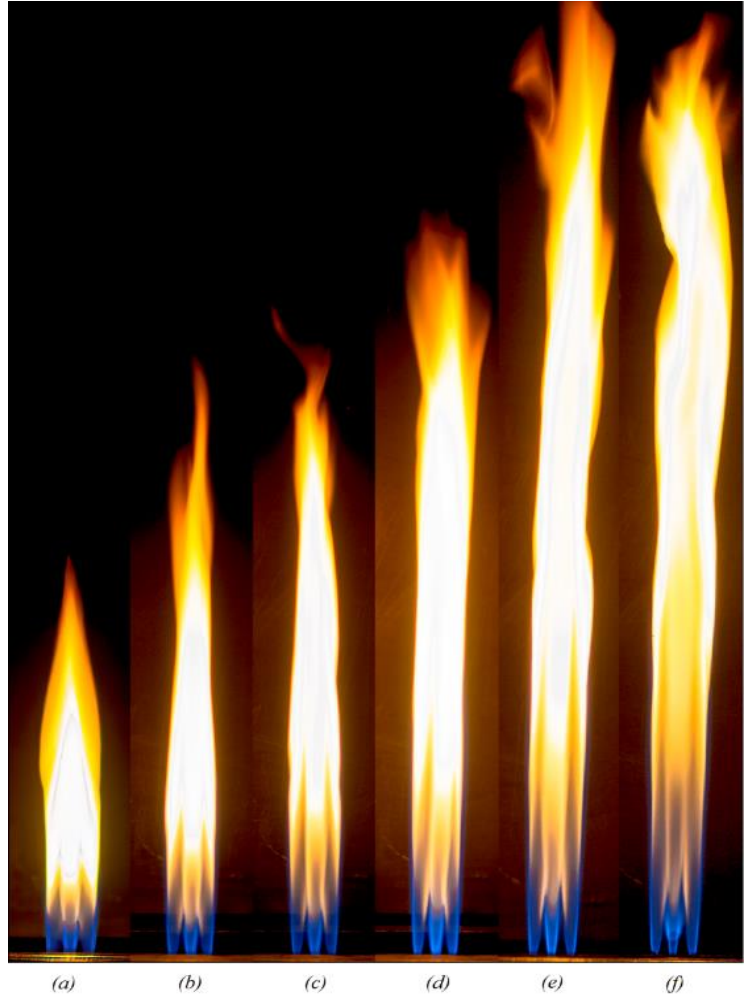

Fig. 6: Overall average flame height as a function of the fuel flow.
(a) 350 cc/min; (b) 650 cc/min; (c) 950 cc/min; (d) 1200 cc/min; (e) 1500 cc/min; (f) 1800 cc/min.

Table 5. Average measured flame heights.

| $\dot{V}_{Fuel}$ | 350 | 650 | 950 | 1200 | 1500 | 1800 |
|---|---|---|---|---|---|---|
| $h_{Flame}$ | 12.85215 | 21.44532 | 27.74474 | 33.17976 | 36.02305 | 38.61717 |

In passing, we note that the flame temperature was discarded from the modelling based on the previously presented low coefficient of variation, together with the fact that when conducting an ANOVA from the measured magnitudes, the temperature presents the highest P-Value with 0.6431, which means that within 95% CI, the term is not statistically significant. Finally, in Figs. 7 to 9 the proposed height correlation is compared to the most accepted ones [32,33,38,66–68] in order to evaluate the adjustment between them and against the measured magnitudes.

It should be noticed that the time variation of the flame height can be regarded as small so that the flame is in a quasi-steady state. Under the previous assumption McCaffrey [66], Heskestad [32,33] and

Alpert and Ward [68] reported that the total flame height can be estimated by an extended correlation model regarding a buoyancy controlled flame height. Moreover, Delichatsios [38] developed a correlation based on an heat-released-rate calculation by considering the buoyancy effects through the Froude number, called *flame Froude number $Fr_f$*. A comparison between all these correlations is given in Table 6.

Table 6: Compared flame height correlations.

| | *Flame type* | *Correlation* | |
|---|---|---|---|
| Proposed Luminous Flame Height Correlation | Total flame height | $h_{Flame} = \left(0.01216097953 + \dfrac{22.82969809}{\dot{V}_{Fuel}}\right)^{-1}$ | (14) |
| McCaffrey | Intermittent buoyant flame length | $h_{Flame} = 0.08\dot{Q}^{2/5}$ | (15) |
| | Continuous flame height | $h_{Flame} = 0.20\dot{Q}^{2/5}$ | (16) |
| Heskestad | Intermittent buoyant flame length | $h_{Flame} \approx 0.07\dot{Q}^{3/5}$ | (17) |
| | Continuous flame height | $h_{Flame} = 0.235\dot{Q}^{2/5} - 1.02D$ | (18) |
| Alpert & Ward | Intermittent buoyant flame length | $h_{Flame} \approx 0.07\dot{Q}^{3/5}$ | (19) |
| | Continuous flame height | $h_{Flame} = 0.174k\dot{Q}^{2/5}$ | (20) |
| $Fr_f$ | Continuous flame height | $h_{Flame} = \dfrac{L^* d_j (\rho_e/\rho_\infty)^{1/2}}{f_s}$ | (21) |
| | Dimensionless flame length $L^*$ | $L^* = \dfrac{13.5 Fr_f^{2/5}}{(1 + 0.07 Fr_f^2)^{1/5}}$ | (22) |
| | Flame Froude number $Fr_f$ | $Fr_f = \dfrac{v_e f_s^{3/2}}{\left(\dfrac{\rho_e}{\rho_\infty}\right)^{1/4} \left[\dfrac{\Delta T_f}{T_\infty} g d_j\right]^{1/2}}$ | (23) |

Here $\dot{Q}$ is the heat release rate in kilowatts, $D$ is the flame equivalent diameter or nozzle equivalent diameter, $k$ is a fire confinement constant, i.e., 1, 2, 3, $d_j$ is the fuel jet diameter, $\rho_e$ is the fuel density, $\rho_\infty$ is the ambient density, $v_e$ is the fuel velocity, $f_s$ is the stoichiometric mixture fraction, $\Delta T_f$ is the temperature flame gradient between the ambient and $T_{adiab}$ (or $T_{measur}$) and $g$ is the gravity. At small values of $Fr_f$ (Fr<5), equation (22) simplifies to $L^* = 13.5 Fr_f^{2/5}$, the buoyancy-dominated limit. As the value of $Fr_f$ increases (Fr>5), the dimensionless flame lengths predicted by this correlation asymptotically approach the momentum-dominated dimensionless flame length value $L^* = 23$.

The existing discrepancy is primarily attributed to the fact that the existing correlations were developed by considering only fuel properties, especially its heat release rate. Secondly, the amount of fuel mass flow remains in a transition-to-turbulent regime. Third, it was probed for hydrogen, methane and distinct concentrations of LPG and *D*. In addition, the visualization methodology proposed in this study does not separate the luminescent and chemical reactions regions, thus a certain inherent difference exists between the optically measured and effective flame heights.

As illustrated in Tables 7 and 8, the discrepancy between the aforementioned correlations and the proposed one is, in general, due to the handling of the fluid regime employed to develop the flame height. The results of the analysis presented entail that the proposed correlation can handle better the laminar regime and the general behaviour resembles the McCaffrey's correlation, albeit its distribution is similar to the Heskestad's [32] and $Fr_f$ correlations. This is mainly because both correlations use the same $\dot{Q}^{2/5}$ and almost the same value of the constant, i.e., 0.2 and 0.235, respectively. On the other hand, the $Fr_f$ correlation handles buoyancy better than McCaffrey's and Heskestad's correlations.

Table 7: Overall adjustment of the proposed model compared to the existing ones.

|  | Distribution Fit % | Overall Average Discrepancy % |
|---|---|---|
| McCaffrey | 0.17 | 1.75 |
| Heskestad | 0.09 | 4.54 |
| Alpert and Ward | 0.33 | 7.4 |
| $Fr_f$ | 0.17 | 4.54 |

Table 8: Overall adjustment of the proposed correlation as a function of the volumetric flow of fuel.

| Fuel cc/min | 350 | 650 | 950 | 1200 | 1500 | 1800 |
|---|---|---|---|---|---|---|
| Overall Average Variation % | 16.9 | 8.81 | 5.7 | 5.15 | 5.43 | 5.89 |

The correlation obtained in this work predicts both the consistent and flickering flame height and leads to the following statements:

- McCaffrey's and Alpert and Ward's correlations evolve as a function of the fluid flow by sharing the same tendency of adjustment of the flame height.
- The Hekestad's correlation matches the proposed correlation more consistently at high fuel flows.
- The Alpert and Ward's correlation has the most extended deviation compared to the proposed correlation.
- The $Fr_f$ correlation does not work, in general, for fuels having different properties.
- The $Fr_f$ correlation for buoyancy-momentum transition characterizes the aerodynamic effects of the nozzle diameter in the development of the buoyant flow.
- All correlations cannot handle laminar flame flickering separately but matches, in a large degree, the transition-to-turbulent and turbulent flame flow asymptotic behaviour.

The flame luminosity in the lower part of the combustion region is usually more stable, while the upper part exhibits vortex structures [66], more or less pronounced, that form near the base of the flame and propagate upward leading to intermittency. This implies that the ratio of released gases influences directly the flame tip in the form of flickering flame kernels with a specific cut-off frequency. In buoyant laminar diffusion flames, the tip low frequency of flickering and necking normally oscillates in a range between 10 and 20Hz, depending upon the operating conditions [69–73]. This is mainly because the laminar flame flickering, which is initially influenced by the entrained airflow and then by the free mixing process, causes that the flame-tip cuts in the form of kernels are not perceived in a sufficiently definite way. Consistent with the above analysis, the McCaffrey correlation, which is based on the flame heat release does not take into account this detached kernels, which were included as averaged flame structures.

Although McCaffrey's, Heskestad's and Alpert and Ward's correlations are supported by the rearrangement of the Froude number to cope with the flickering of the flame tip, they are designed to deal with a high-momentum jet flame regime, where the flame height ceases to vary with the fuel flow rate and is several hundred times the size of the source diameter. Chen *et al*. [71] found that the flame bindings related to its flickering behaviour were conserved through the mixing and developed flame structure until the end of the mixing process in a transition to a turbulent flame-species mixing process. Furthermore, the turbulent flame tip has a distinct flickering frequency as a function of the increment of species mass flow. In line with these findings, the flame height in the near the most transition-to-turbulent regime, the detached flame kernels occur in a high frequency range from 130 to above 500Hz [8,74]. Thus, when the fuel mass flow increases the high frequency at the flame tip appears to be more stable and flame-tip cuts seem to fade.

Contrary to what is observed in Fig. 6, these flame-tip cuts remain bound to the combustion products, which behave as gases that absorb the radiation from the combustion, thereby causing the flame-tip luminosity to be deficient due to its blurred nature for the image-capture instrument. The temperatures of the combustion products could be captured by other means, i.e., by thermography or by infrared radiation [52,75–78]. However, this analysis is far beyond the scope of this paper. All the above correlations point to the diffusion process as a driver of the consistent flame length, which is integrated by the consistent luminous flame height, the luminous flame of detached low-frequency kernels and the high-temperature combustion products flame length. This last length would confirm the height of the chemical flame that Hottel [50] describes with the relation of the CO and $CO_2$ ratio. This length is also captured by the cameras but fuses behind the luminous zone of the still hot combustion products and high-frequency flame-kernels cuts. The flame length discrepancies shown in Figs. 7 to 9 might be attributed to slight inconsistencies between the cameras. Nonetheless, through this analysis it can be corroborated that the direction of the camera does not significantly affect the measurements, these being only attributable to the development of the flame itself.

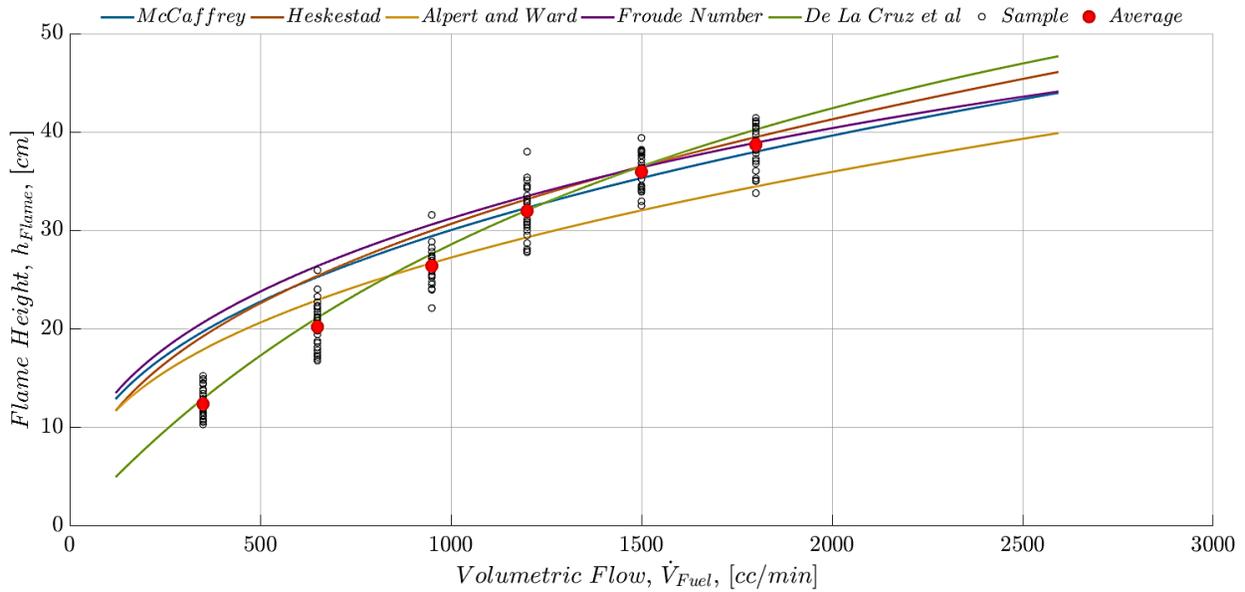

Fig. 7: Samples and average flame height, $h_{Flame}$, from a lateral view.

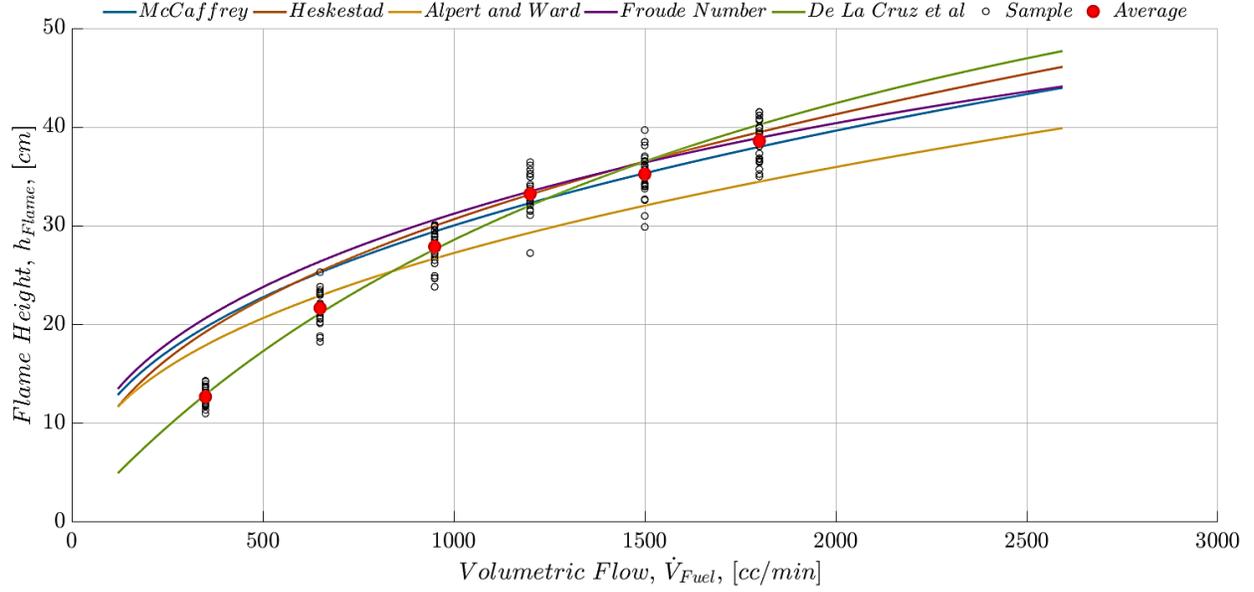

Fig. 8: Samples and average flame height, $h_{Flame}$, from a 45° angle view.

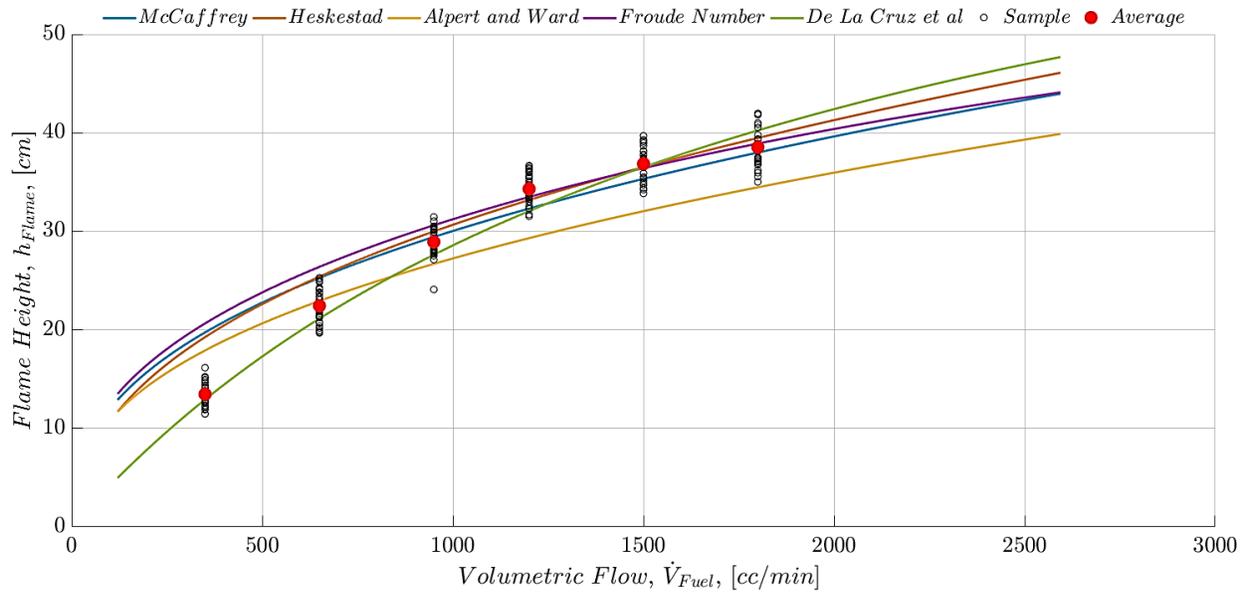

Fig. 9: Samples and average flame height, $h_{Flame}$, from a frontal view.

### 4.4 UNCERTAINTY MEASUREMENT

Based on the sample size and the information obtained from the experimental dataset, and through both the Kolmogorov-Smirnov and Kruskal-Wallis tests, it is concluded that within twice the standard deviation, the distribution function of the measurements roughly approximates a normal distribution. Therefore, a type A uncertainty measurement is obtained [79]. The type A uncertainty is characterized by the observed frequency distribution and statistical analysis of the measured quantity, as a means to evaluate the uncertainty [80]. The results are given in Tables 9 and 10.

Table 9: Main statistics and standard error of the measurements.

| Camera Direction | Fuel Flow | Mean | Std Dev | Variance | Df | SE | SS | RSS |
|---|---|---|---|---|---|---|---|---|
| Lateral | 350 | 12.38939 | 1.45622 | 2.12057 | 31 | 0.26154 | | |
| | 650 | 20.20944 | 2.213045 | 4.89756 | 31 | 0.39747 | | |
| | 950 | 26.40976 | 1.653782 | 2.73499 | 31 | 0.29702 | | |
| | 1200 | 31.98872 | 2.260229 | 5.10863 | 31 | 0.40594 | | |
| | 1500 | 35.96005 | 1.661384 | 2.76019 | 31 | 0.29839 | | |
| | 1800 | 38.70572 | 2.059402 | 4.24113 | 31 | 0.36987 | | |
| Angled | 350 | 13.46626 | 1.075218 | 1.15609 | 31 | 0.19311 | | |
| | 650 | 22.44556 | 1.639631 | 2.68839 | 31 | 0.29448 | | |
| | 950 | 28.92478 | 1.458384 | 2.12688 | 31 | 0.26193 | 5.36525 | 2.3163 |
| | 1200 | 34.30298 | 1.457174 | 2.12335 | 31 | 0.26171 | | |
| | 1500 | 36.86413 | 1.574273 | 2.47833 | 31 | 0.28274 | | |
| | 1800 | 38.54574 | 1.81171 | 3.28229 | 31 | 0.32539 | | |
| Frontal | 350 | 12.70079 | 0.835059 | 0.69732 | 31 | 0.14998 | | |
| | 650 | 21.68097 | 1.525576 | 2.32738 | 31 | 0.27400 | | |
| | 950 | 27.89969 | 1.638748 | 2.68549 | 31 | 0.29432 | | |
| | 1200 | 33.24758 | 1.707904 | 2.91693 | 31 | 0.30674 | | |
| | 1500 | 35.24496 | 2.037165 | 4.15004 | 31 | 0.36588 | | |
| | 1800 | 38.60006 | 1.807535 | 3.26718 | 31 | 0.32464 | | |

Table 10: Type A uncertainty measurement.

| Statistic | Value |
|---|---|
| Pooled Variance | 2.875712 |
| Pooled Standard Deviation | 1.695793 |
| Standard Uncertainty | 0.304573 |
| Coverage Factor | 2 |
| Expanded Uncertainty | 0.609147 |

From Tables 9 and 10 it follows from the entire experimental dataset that the root sum squared of the standard error (RSS) is equal to 2.31cm for each treatment, accounting for the variability between the samples. In order to compute the associated uncertainty, the method of pooled variance, $\sigma_p^2$, is employed and from this, the standard uncertainty, $u(y_i)$, is obtained to be 0.304573 cm. Consequently, to keep such an uncertainty within the 95% CI, a coverage factor $k = 2$ is assigned [80], obtaining an expanded uncertainty, $U$, with a value of 0.609147cm, which is expressed as an uncertainty measure for the flame height equivalent to ±11%.

### 4.5 NUMERICAL SIMULATIONS ANALYSIS

Because, the flame height is also conformed by chemical flame length, the further section describe the relevance of the chemical flame height by means of numerical simulation. Its contribution to the flame evolution is also presented to take into account the development of the correlation.

One way to separate the flame length into luminous and chemical flame heights is to trace the maximum flame temperature, called the core flame temperature. By doing so, it is well known that $CO_2$ is the last product species formed, which indicates that oxygen radicals are completely consumed. Hence, according to Hottel [50], the chemical flame height takes relevance when the concentration ratio $CO/CO_2$ is up to 0.15. Krishnamoorthy and Ditaranto [81] argued that the chemical flame length can be obtained by estimating the flame shape as the locus of points where the CO mol fraction is reduced to 1%. Assuming this last value and Hottel's ratio, an upper limit of 0.07 is obtained for the $CO_2$ mole fraction. On the other hand, Yagi [19] establishes a peak flame temperature of 1473°C where the $CO_2$ mol fraction measured was on its maximum value. Using this last argument, for this temperature value, the $CO_2$ mol fraction has a value of 0.02. This range establishes a length located between the luminous and flickering flame zone as shown in Fig. 10. With the aid of the numerical simulations, it was possible to rebuild the total flame height because the chemical length must also be taken into account as is indeed captured by the cameras. This length was also accounted for in the development of the correlation given by equation (14).

Figure 10 shows the differences between the experimental and simulation datasets. For the whole set in this evaluation, the sum of the forecasts comes to 160.76, which is 9.10 lower than the sum of the observations. Hence, the forecasts are biased 1.51 degrees down. Of the six forecasts, only case one had a forecast higher than the observation. Therefore, it is inferred that there is an underlying reason causing the forecasts to be low. This is attributed to the fact that the visualization methodology employed cannot separate the luminescent and the chemical reactions, thus an intrinsic increment is associated with the measuring process.

However, there are no large errors in this case; the highest one corresponding to a 3.26 degree error for case five. Consequently, counting the squares of the errors amounts to 32, leading to a RMSE of 2.30, which is not much higher than the bias of 1.51. This implies that a significant part of the error in the forecasts is due solely to the persistent bias. Hence, to minimize the RMSE it is imperative to try different visualization techniques in order to mitigate the aforementioned cause, thereby reducing the bias.

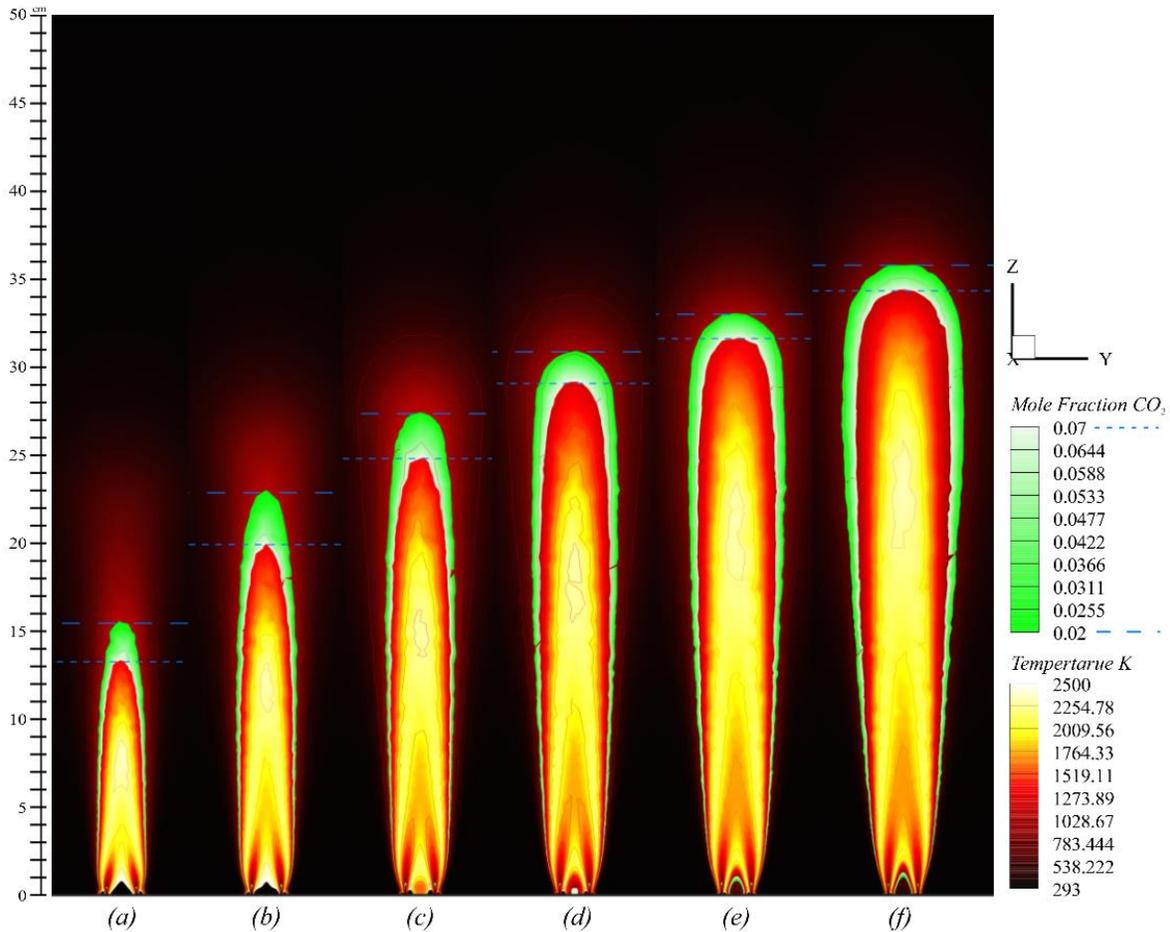

Fig. 10: Chemical flame heights based on molar $CO_2$ concentration as a function of temperature.

In addition, Fig. 11 shows the relative error for each case, where the highest one corresponds to case one (i.e., the 350cc/min flow) with a 12.04% and the lowest error is found for case two, corresponding to the 650cc/min flow, with 0.25%. The remaining cases corresponding to 950cc/min, 1200cc/min, 1500cc/min and 1800cc/min flows have relative errors of 5.83%, 8.3%, 9.05% and 7.65%, respectively. This discrepancy can be attributed mostly to the previously computed uncertainty, which is particularly close to the maximum relative error.

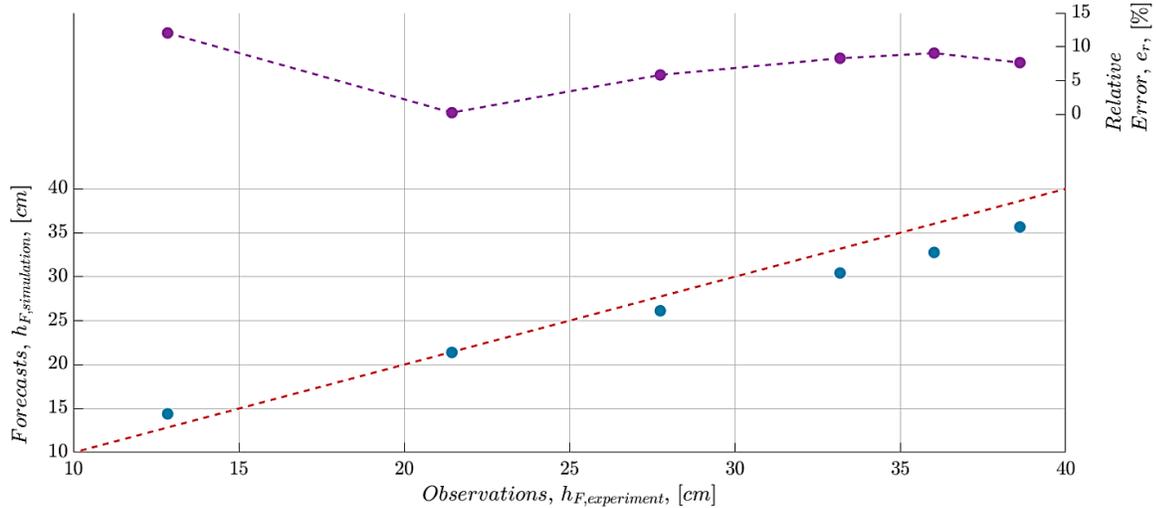

Fig. 11: Scatter plot of the adjustment between the experimental and simulation datasets.

## 5 CONCLUSIONS

A flame height correlation for laminar to transition-to-turbulent diffusion flames was established, which is given by the relation

$$h_{Flame} = \left(0.012161 + \frac{22.8297}{\dot{V}_{Fuel}}\right)^{-1}. \tag{14}$$

The experiments show that this correlation has an adjustment of luminous flame height in the laminar regime of 16.9%. This indicates that without the use of the intermittent buoyant flame height correlation, this correlation provides a better representation of the flame height in this regime. Compared to most accepted flame height correlations, the adjustment variation is on average 5.54% for the transition-to-turbulent regime, which means that it fits as much as the aforementioned correlations at these regimes.

The general behaviour of the proposed correlation resembles that of McCaffrey's correlation, albeit its distribution is similar to Heskestad's and $Fr_f$ correlations. For example, Hekestad's correlation matches this correlation only at fuel flows with Re>1351. All correlations cannot handle laminar flame flickering separately. However, with the exception of Alpert and Ward's correlation, they match the laminar and the transition-to-turbulent flame flow asymptotic behaviour. Moreover, Alpert and Ward's correlation exhibits the greatest average deviation (of 7.4%) compared to the proposed correlation.

The numerical results show that the predicted range for the chemical flame length is located between the luminous and flickering flame zone as compared with the experimental flame images. These results agree with the predictions of the chemical length zone found in the literature. All this numerical information complements the data necessary to fit the chemical, the buoyant and the flickering flame length for the development of the proposed correlation. Therefore, this proposed correlation, only based on the mass fuel flow can be used from laminar to transition-to-turbulent combustion regimes for the port array burner configuration.

## 6 ACKNOWLEDGMENTS

The authors acknowledge the support provided by the National Autonomous University of Mexico Engineering Institute and grants provided by the National Council of Science and Technology of Mexico (CONACyT) and the Mexican Ministry of Energy (SENER) as well as the support provided by Laboratory of Applied Thermal and Hydraulic Engineering, Superior School of Mechanical and Electrical Engineering of the National Polytechnic Institute is also acknowledged. Likewise, the support provided to the National Institute for Nuclear Research is appreciated (ININ).

## 7 FUNDING

The research work described in this paper was fully supported by the grants from the National Council of Science and Technology of Mexico (CONACyT) and the Mexican Ministry of Energy (SENER), as well as the resources provided by the Applied Thermal and Hydraulic Engineering Laboratory of the National Polytechnic Institute of Mexico.

## 8 AUTHORS CONTRIBUITIONS

**M. De La Cruz-Ávila**: Conceptualization, Investigation, Software, Data Curation, Resources, Visualization, Writing – Original Draft, Review & Editing, Project Administration, Funding Acquisition;
**J.E. De León-Ru**iz: Conceptualization, Methodology, Software, Validation, Formal Analysis, Investigation, Data Curation, Visualization, Writing – Original Draft, Review & Editing;
**I. Carvajal-Mariscal**: Methodology, Software, Validation, Formal Analysis, Investigation, Resources, Writing – Review & Editing, Supervision, Project Administration;
**G. Polupan**: Methodology, Validation, Visualization, Writing – Review & Editing;
**L. Di G. Sigalotti**: Validation, Writing – Review & Editing.